\documentclass[12pt]{article} 
\usepackage[koi8-r]{inputenc}
\usepackage{cyr}
\usepackage{epsf}
\usepackage{pazha}
\usepackage[round]{natbib}
\usepackage{graphicx}
\usepackage{hyperref}
\tightenlines

\usepackage{amsmath} 
\usepackage{xcolor}

\sloppypar

\begin{document}
\baselineskip 21pt

\title{\textbf{SIMULATION OF A PROTOPLANETARY DISK ACCRETION ACTIVITY DUE TO A COLLISION WITH A GAS STREAM}}

\author{\textbf{\hspace{-1.3cm}\copyright\, 2025. \ \ 
V.~V.~Grigoryev$^{1*}$, T.~V.~Demidova$^1$}}

\affil{
{\textit{$^1$Crimean Astrophysical Observatory of the Russian Academy of Sciences, Nauchny, Crimea, Russia}}}

\vspace{2mm}
\received{30.07.2025 \\
After revision 03.09.2025; accepted for publication 03.09.2025}
\sloppypar 
\vspace{2mm}
\noindent
The consequences of a protoplanetary disk collision with a gas stream are being studied using three-dimensional numerical gas-dynamic simulation. The influence of orbital parameters and the stream mass on the accretion activity of the star is examined. It is shown that the orbital inclination and the initial mass of the infalling material are the most influential parameters in determining the accretion rate. The obtained accretion rate dependencies are compared with actual observational data for two FU~Ori type stars. It turns out that not only is the maximum accretion rate consistent with observational estimates, but the behavior of the accretion rate over time is very similar to available long-term light curves.
\noindent


\vfill
\noindent\rule{8cm}{1pt}\\
{$^*$ email: vitaliygrigoryev@crao.ru}

\clearpage

\section*{INTRODUCTION}
\noindent
One of the first hypotheses about the causes of powerful outbursts of luminosity in objects like FU Ori was the idea that a sharp increase in accretion onto the star was caused by the fall of a large clump of matter from the remains of a protostellar cloud onto the circumstellar disk~\citep{1985ApJ...299..462H, 1996ARA&A..34..207H}.

Observations of protostellar accretion disks have revealed that small filaments of matter, denser than their surroundings, known as streamers, exist in their vicinity~\citep{2022A&A...667A..12V,2023ApJ...958...98F,2023A&A...669A.137H}. Their matter settles on disks at specific points. They also heat the disk through accretion shocks and locally increase the disk mass, which, in turn, can cause gravitational instability. Similar structures have also been detected in simulations of the formation of protostellar accretion disks~\citep{2024A&A...686A.246H,2025MNRAS.543.3321M}.

The flow of material accreting onto the FU Ori system was detected using the ALMA radio interferometer and described in~\citet{2024ApJ...966...96H}. Although the accretion rate is significantly lower than that observed in young stars, the authors believe it may be the remnant of a more powerful stream that triggered the burst of accretion activity in FU Ori.

\citet{2018MNRAS.475.2642K} showed that dense clumps of matter form in the vicinity of the star during the formation of the circumstellar disk. These clumps subsequently fall onto the inner regions of the circumstellar disk (several AU from the star). Episodes of such delayed accretion were detected for all models calculated by the authors.

The first gas-dynamic simulation of the infall of a clump of matter in the immediate vicinity of a star was performed by~\citep{2023ApJ...953...38D} in the approximation of discrete addition of several portions of matter to the protoplanetary disk. The calculations were performed using the smoothed particled hydrodynamic method (SPH). According to the obtained results, for an accretion rate burst to resemble an FU Ori-type flare in amplitude and timing, the perturbation must be located closer than 3 AU, its mass must be at least $3M_J$ ($M_J$ is the mass of Jupiter), and the velocity of the matter must not exceed $60\%$ of the local Keplerian velocity. 

\citet{2023PCT...1868..269G} compared the results of calculations for the decay of a dense clump in a rotating gas disk using two methods: the SPH method and the finite volume method. Evidence was obtained that the results are qualitatively consistent.

The next step was to model the free fall of a gas stream onto a protoplanetary disk~\citep{2024Proc...40G,2024ARep...68..949G}. This showed that the greatest loss of kinetic energy from the stream occurs in the case of a retrograde (relative to disk rotation) collision with the gas disk, in which the initial orbit of the stream intersects the plane of the disk twice. Moreover, for the accretion rate onto the star to sharply increase by two orders of magnitude, it is sufficient for the initial mass of the clump to be equal to one mass of Jupiter. These studies investigated a collision between the stream and the disk at a distance of $\sim 7$~\text{AU}.

In this paper, we examine the consequences of a gas stream's passage through protoplanetary disk material for a wide range of initial stream orbit parameters and several initial masses. We demonstrate and explore the possibility of a sharp flare in accretion rate, consistent with FU Ori-type objects, in terms of shape and amplitude.

\section*{MODEL AND METHOD}
\subsection*{Statement of the problem}
It is assumed that a solar-mass star is surrounded by a gaseous protoplanetary disk. A dense clump of matter falls freely onto the disk.

At the initial moment, the disk's volume density is distributed according to the law
\begin{gather}\label{eq:rho}
	\rho(r, z, 0) = \rho_0\left(\frac{r}{r_{in}}\right)^p\exp\left(\frac{GM}{c_s^2(r)}\left[\frac{1}{\sqrt{r^2+z^2}}-\frac{1}{r} \right]\right).
\end{gather}
Here $r = R\sin \theta$~is cylindrical radius ($R$~is spherical radius, $\theta$~is polar angle), $z$~is altitude above the equator, $c_s(r)$~is the speed of sound at the radius, $r_{\textrm{in}}$~is inner radius of the disk, $G$~is gravitational constant, $M$~is stellar mass.

The speed of sound in the disk was determined as follows: $c_s^2(r)=c_0^2\left(\frac{r}{r_{\textrm{in}}}\right)^q$, where $c_0$~is speed of sound at a distance $r_{\textrm{in}}$. The speed of sound together with the angular Keplerian velocity $\Omega_K(r)=\sqrt{GM/r^3}$ determines the half-thickness of the disk $H(r)=c_s(r)/\Omega_K(r)=H_0\Big(\frac{r}{r_{\textrm{in}}}\Big)^{(q+3)/2}$, where $H_0=c_0/\Omega_K(r_{\textrm{in}})$.

The initial gas distribution parameters were chosen in accordance with empirical laws derived from observations of protoplanetary disks. Specifically, the values $p = -2.25$ and $q = -0.5$ were adopted. For a given $p$, the vertical density distribution is determined by the condition of hydrostatic equilibrium. The surface density in this case is determined by the formula $\Sigma\approx\Sigma_0\left(\frac{r}{r_{\textrm{in}}}\right)^{-1}$, predicted by~\citet{1997ApJ...486..372B} for accretion disks. According to observational data in the submillimeter range, the $\Sigma$ profile is described as follows: $\Sigma \propto R^{-0.9}$~\citep{2009ApJ...700.1502A}. \citet{1987ApJ...323..714K} showed that the temperature profile of a disk whose thickness increases with radius can be described by a power law $T \propto R^q$, where  $q=-0.5$. This profile describes well most of the observed spectral energy distributions (SEDs) of young stars with disks~\citep{1997ApJ...490..368C}. 

The value of $\Sigma_0$ is determined by the total mass of the disk, which in the paper for all calculations is equal to $M_d=0.01M_\odot$~\citep{2011ARA&A..49...67W}. At the same time $\rho_0=\frac{\Sigma_0}{\sqrt{2\pi}H_0}$. The value of $c_0$ was chosen so that the temperature at the inner boundary of the disk  ($r_{\textrm{in}}=0.2$~\text{AU}) was equal to $T_0=\frac{c_0^2\mu m_H}{\gamma k_B}=1000$~K, in that case $H_0\approx 0.006$~AU. The molar weight of the gas in the disk is $\mu = 2.35$~\citep{1994A&A...286..149D}, $m_H$~is mass of a hydrogen atom, and $k_B$~is Boltzmann constant.

The density in the calculations is limited from below by the value $10^{-12} \rho_0$, and the entire region initially occupied by such a rarefied medium is considered a corona with a temperature of $10^4$~K and a molar weight of $\mu=0.71$.

The initial angular velocity of the matter is given by the formula~\citep{2013MNRAS.435.2610N}:
\begin{gather}\label{eq:Om}
\Omega(r,z)=\Omega_K(r)\left[(p+q)\left(\frac{H}{r}\right)^2+(1+q)-\frac{qr}{\sqrt{r^2+z^2}}\right]^{1/2}.
\end{gather}
Thus, the velocity components at each point on the disk in spherical coordinates are given by
$\vec{v} = (v_R, v_{\theta}, v_{\varphi}) = (0, 0, -\Omega r)$,  i.e., the disk initially rotates clockwise in the $xy$ plane. In previous studies, we have shown that the most powerful accretion bursts will be observed during retrograde infall.

Similar initial conditions were considered in~\citet{2024AstL...50..625D} in a model of a gas disk collision with a free planet. Calculations showed that over a 750-year time interval, despite the planet's gravitational influence, the rate of accretion of disk matter onto the star showed no significant changes. In a previous study~\citep{2024ARep...68..949G} on the prograde infall of a gas stream toward the disk during a single intersection, it was found that the accretion rate became stationary and small (on the order of $10^{-7}$ solar masses per year) approximately 140 years after the start of the calculations. 

At the initial moment, a clump of matter is added to the system: a region of size $\sim 0.8$~AU at a distance $R = R_0$ from the star is filled with matter with a mass $\ensuremath{M_{c}}$ (one computational cell at ``single'' resolution). The initial temperature of this matter is $50$~K. The initial components of the gas velocity in a given cell are specified such that a material point with coordinates at the center of this cell moves along a parabolic trajectory with the longitude of the ascending node $\Omega = 0\ensuremath{^\circ}$ relative to the star. Several motion options were considered, as well as different initial masses in the cell. The calculation parameters are listed in Table~\ref{tab:models}. 
\newpage
\begin{table}[h]\centering	 
	\caption{Calculation parameters  }
	\label{tab:models}
	\begin{tabular}{lcccccc}
		Name  & $q$ & $i, \ensuremath{^\circ}$ & $\omega, \ensuremath{^\circ}$ & \ensuremath{M_{c}} &   Nx    & $R_0$\\ \hline
		q5i15w90  &  5  & 15  &    90    &         1.0         &       1x        & 20\\
		q5i30w90  &  5  & 30  &    90    &         1.0         &       1x        & 20\\
		q5i45w90  &  5  & 45  &    90    &         1.0         &       1x        & 20\\       
		q5i60w90  &  5  & 60  &    90    &         1.0         &       1x        & 20\\
		q5i30w15  &  5  & 30  &    15    &         1.0         &       1x        & 20\\
		q5i30w45  &  5  & 30  &    45    &         1.0         &       1x        & 20\\
		q5i30w135 &  5  & 30  &   135    &         1.0         &       1x        & 20\\
		q5i30w165 &  5  & 30  &   165    &         1.0         &       1x        & 20\\
		q5i30w90  &  5  & 30  &    90    &         0.3         &       1x        & 20\\
		q5i45w90  &  5  & 45  &    90    &         0.3         &       1x        & 20\\
		q5i60w90  &  5  & 60  &    90    &         0.3         &       1x        & 20\\
		q5i30w90  &  5  & 30  &    90    &         0.1         &       1x        & 20\\
		q10i30w90 & 10  & 30  &    90    &         1.0         &      1.5x       & 30\\
		q15i30w90 & 15  & 30  &    90    &         1.0         &       2x        & 40\\
		q5i30w90  &  5  & 30  &    90    &         1.0         &       2x        & 20\\ q10i45w90 & 10  & 45  &    90    &         1.0         &      1.5x       & 30\\
		q10i45w45 & 10  & 45  &    45    &         1.0         &      1.5x       & 30\\ \hline
	\end{tabular}
	
{\textbf{Note.}} $q$~is pericentric distance of the initial orbit (in \text{AU}), 
$i$~is the inclination of the initial orbit of the stream to the plane of the disk (also known as $xy$ plane), $\omega$~is argument of periapsis of the initial orbit, \ensuremath{M_{c}}~is mass of the clump (in Jupiter masses), 
Nx~is resolution (single, one and a half or double), $R_0$~is initial distance (in \text{AU}).
\end{table}
%\newpage

\subsection*{Basic equations}
Using the PLUTO package\footnote{\href{http://plutocode.ph.unito.it/}{http://plutocode.ph.unito.it/}} \citep{2007ApJS..170..228M}, a system of non-stationary gas-dynamic equations was solved in a spherical coordinate system~\citep[similar to the paper of][]{2024ARep...68..949G}:

\begin{equation} \label{eq:main}
 \begin{cases}
		\displaystyle \frac{\partial \rho}{\partial t} + \nabla \cdot \left( \rho \mathbf{v} \right)  =  0\\
 \displaystyle\frac{\partial (\rho \mathbf{v})}{\partial t} + \nabla \cdot \left( \rho \mathbf{v} \cdot \mathbf{v} -  p \hat{I} \right)^T =  - \rho \nabla \Phi + \nabla \cdot {\sf \Pi}(\nu),\\	
	\displaystyle\frac{\partial (\varepsilon_t + \rho \Phi) }{\partial t} + \nabla \cdot \left[ \left(\varepsilon_t + p + \rho \Phi \right)\mathbf{v}   \right]  = \nabla \cdot (\mathbf{v} \cdot {\sf \Pi}(\nu)) + \nabla \cdot \mathbf{F}_c.\\
   p = \rho \epsilon (\gamma - 1),~ \gamma = 1.4. 	
  \end{cases}
\end{equation}

where the density of the gas is indicated $\rho$, velocity is $\vec{v}$, pressure is $p$, $\hat{I}$~is identity matrix, index $T$~is transposition operation, $\Phi = -GM_\ast/R$~is gravitational potential created by a star ($M_\ast = 1 M_{\odot}$~is stellar mass, $R$~is distance to the star, $M_{\odot}$~is Solar mass). ${\sf{\Pi}}(\nu)$~is viscous stress tensor:	${\sf \Pi}(\nu) = \nu_1 \left[ \nabla \vec{v} + (\nabla \vec{v})^T\right] + \left(\nu_2 - \frac{2}{3}\nu_1 \right)(\nabla \cdot \vec{v})\hat{I}$, where $\nu_1$~is coefficient of kinematic viscosity, $\nu_2$~is second viscosity ($\nu_2=0$). The total energy density $\varepsilon_t = \rho \epsilon + \rho \mathbf{v}^2/2$ includes the specific internal energy of the gas $\epsilon$. The heat flux is determined by the thermal conductivity coefficient $\kappa$ and the temperature gradient of the gas $T$: $\mathbf{F}_c = \kappa \cdot \nabla T$.

The constant adiabatic index $\gamma$ was chosen equal to $1.4$ because most of the studied material is molecular hydrogen. Initially, the maximum temperature in the disk (near the inner boundary) is $T_0 = 1000$~K, a value insufficient for dissociated molecules to significantly contribute to the change in the adiabatic index. Furthermore, accounting for the change in $\gamma$ in the calculations requires solving additional equations (for example, the Saha equation), which complicates the calculations but does not qualitatively change the picture due to the lack of consideration of radiative transfer and the magnetic field.

In the hot, rarefied corona, laminar ionic viscosity and electron thermal conductivity dominate, while in the cold, dense disk, turbulent viscosity and thermal conductivity dominate. Therefore, it is necessary to create a sufficiently smooth and physical representation of the corresponding coefficients, depending on the existing gas macroparameters in each individual cell. Molecular viscosity and thermal conductivity make a significant contribution only in the transition layer between the corona and the disk. However, since the disk undergoes vigorous mixing and heating of matter, one can expect, at certain moments in time, the creation of conditions similar to those initially present in the region between the disk and the corona. Therefore, the coefficients of viscosity and thermal conductivity are composed of three components: for neutral 
($\nu_{neutral}$, $\kappa_{neutral}$) 
and ionized ($\nu_{ion}$, $\kappa_{el}$) hydrogen according to \citet{Braginskii1965en} formulas, as well as the turbulent  
term ($\nu_{turb}$, $\kappa_{turb}$).

They are defined by the following formulas: $\nu_{neutral} = 1.016 \times \frac{5}{16 d^2} \sqrt{\frac{2 m_H k T}{\pi}}$, where $d = 2.9 \times 10^{-8}$~cm (diameter of a hydrogen molecule), $\kappa_{neutral} = \frac{1}{3} \rho c_v \lambda v_{therm}$, where $c_v$~is specific heat capacity of a gas at constant volume, $\lambda$~is mean free path of a molecule (but not more than the size of the calculation cell), $v_{therm}$~is thermal velocity of gas (hard sphere model used). $\nu_{ion} = 0.96 \frac{k_b T}{\rho} \tau_i$, where $\tau_i$~is characteristic time of collisions between ions in a fully ionized plasma; $\kappa_{el} = 2 \times 10^{-8} T^{5/2}$ and $\nu_{turb} = \alpha \rho \frac{c_s^2}{\Omega}$,  
where $\alpha = 0.001$ \citep{ShakuraSunyaev1973}, $\kappa_{turb} = c_p \nu_{turb}$, where $c_p$~is specific heat capacity of gas at constant pressure. 

The final viscosity and thermal conductivity are calculated as follows:
\begin{gather}
	\rho \nu_1 = (1 - x_{HI})\mu_{\textrm{ion}} + x_{HI}\mu_{\textrm{neutral}} + \mu_{\textrm{turb}}; \qquad \kappa = (1 - x_{HI})\kappa_{\textrm{el}} + x_{HI}\kappa_{\textrm{neutral}} + \kappa_{\textrm{turb}},
\end{gather}
where $x_{HI}$~is the fraction of neutral hydrogen, which was calculated based on the equilibrium of the processes of ionization and recombination of hydrogen in the cell.

The calculations were performed in the domain $R\in[0.2; 107.2]$~\text{AU}, $\theta\in[15; 165]\ensuremath{^\circ}$, $\varphi\in[0; 360)\ensuremath{^\circ}$ in ``single'' resolution for $144 \times 60 \times 144$ cells, ``one and a half'' for $216 \times 90 \times 216$ cells  and ``double'' one for $288 \times 120 \times 288$ cells. The boundary conditions of the problem are as follows: on the left boundary on $R$, thermal and viscous turbulent fluxes, as well as entropy~\citep[corresponding to][]{2024ARep...68..949G}, are preserved along the radius; on the right, there is a free boundary. Free boundaries are established along the angle $\theta$, and periodic boundaries along $\varphi$.

All macroscopic parameters of the gas are specified in code units. The main parameters are listed in the Table~\ref{tab:dimensionalization}, these designations will also be used further in the text of the paper.

\begin{table}[h]\vspace{6mm}
	\centering
	\caption{Code units} \label{tab:dimensionalization}
	\vspace{5mm}\begin{tabular}{|l|c|c|c|l|}
		\hline
		Parameter           & Designation & Value               & in cgs units  & Comment     \\[2pt] \hline
		Unit of mass    & $M_0$       & $1.98 \times 10^{33}$  & g          & $M_{\odot}$     \\
		Unit of length     & $L_0$       & $1.496 \times 10^{13}$ & cm         & $1$~\text{AU}        \\
		Unit of time   & $t_0$       & $3.16 \times 10^7$     & s       & $1$~yr         \\
		Unit of density & $\rho_0$    & $5.94 \times 10^{-7}$  & g~cm$^{-3}$ & $M_0/L_0^3$     \\
		Unit of velocity  & $v_0$       & $4.74 \times 10^{5}$   & cm~s$^{-1}$  & $2 \pi L_0/t_0$ \\
		\hline
	\end{tabular}
\end{table}

The influence of processes associated with radiation transfer (heating by UV radiation from hot regions of the star, cooling of dust) and with the interaction of dust and gas, as well as the self-gravity of the disk and the magnetic field, on the gas were not taken into account in the paper.

\subsection*{Models}
To study the influence of the clump's orbital parameters on the accretion rate, various trajectories were chosen.
The initial clump mass was also varied. All initial orbital characteristics are presented in Table.~\ref{tab:models}. The following configuration was chosen as the reference calculation against which the comparison will be made: pericentric distance $q = 5~\text{AU}$, the inclination of the clump orbit plane to the disk plane (also known as $xy$ plane) $i = 30\ensuremath{^\circ}$, argument of periapsis of the orbit $\omega = 90\ensuremath{^\circ}$, initial mass of the clump $\ensuremath{M_{c}} = 1.0M_J$, single resolution. 

Typically, the clump was placed at a distance of $R_0 = 20$~\text{AU}~in the half-space $z<0$ and always had the same density and dimensions. Therefore, for calculations with higher resolution (and a larger pericentric distance), the clump was positioned further from the star. The exception was a calculation with double resolution, but placing the clump at a distance of $R_0 = 20$~\text{AU}, which resulted in an initial density eight times higher and linear dimensions are two times smaller.

\section*{RESULTS}
\noindent

\begin{figure}[h!]\centering
	\includegraphics[width=1.0\linewidth]{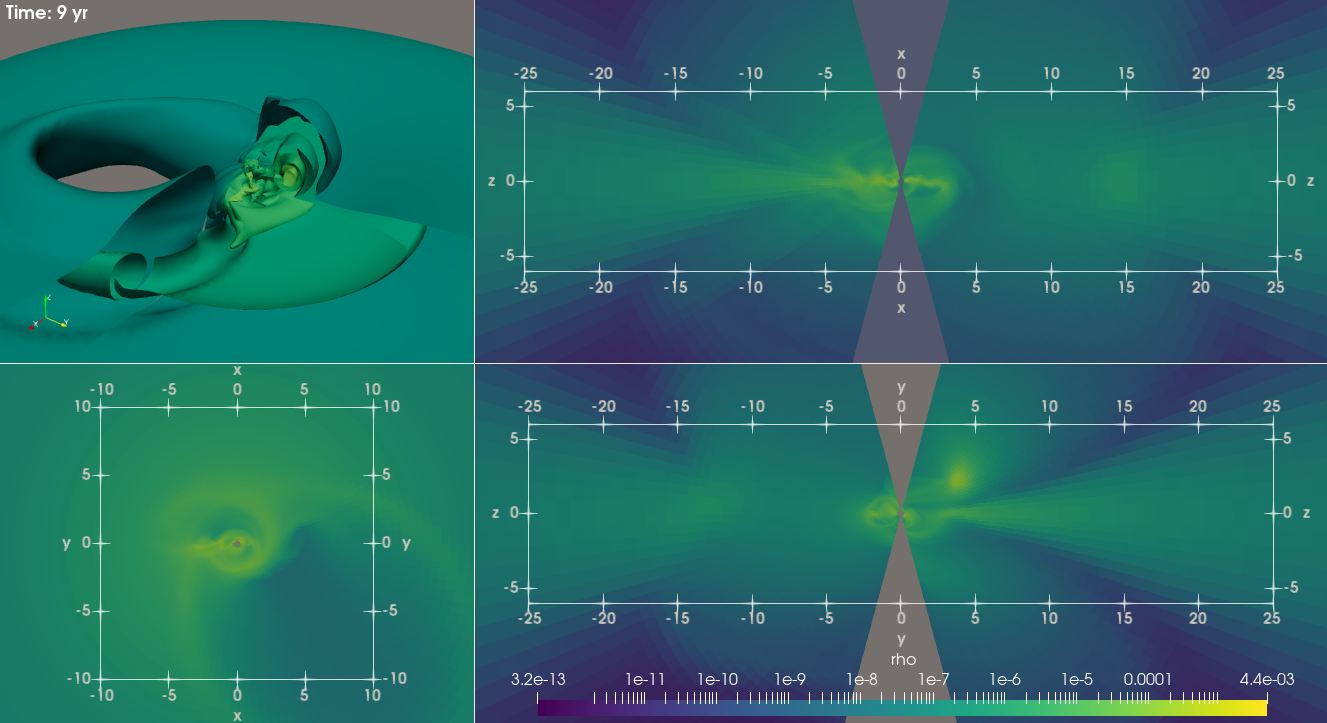}
	\caption{Gas density distribution in the reference calculation at time $9t_0$, close to the moment the clump passes the orbital pericenter. The density is shown in the form of contour surfaces at the top left (the outer surface corresponds to $\rho = 1.7 \times 10^{-4}\rho_0$, the first octet is cut out for clarity of visualization), at the top right is a slice in the $xz$ plane, at the bottom left is a slice in the $xy$ plane, and at the bottom right is a slice in the $yz$ plane. The last image shows the clump itself, and also shows the density scale in $\rho_0$ units, common to all images. Distances along the axes are given in AU.}
	\label{fig:q5i30w90_9} 
\end{figure}

As it falls onto the protoplanetary disk, the clump is stretched in the direction of its motion and, at the moment it crosses the disk plane, forms a stream of finite size. Figure~\ref{fig:q5i30w90_9} shows the density distribution in the reference calculation at approximately the moment the clump passes its pericenter. The volumetric image clearly shows a large ``hole'', which is a region of greatly reduced density in result from the clump's passage through the disk plane. Furthermore, in the half-space $z>0$ near the star, a ``warp'' is present, the clump has stretched
into a stream and risen above the disk plane, passing the pericenter. The dense portion of the stream, surrounded by a less dense medium, is also visible.

In the $xz$ slice, it is clearly seen that in the half-space $x>0$, the disk structure is very different (in fact, this is the region of the hole) from that in the half-space $x<0$, where the density distribution is still close to the initial hydrostatic one. In addition, gas streams elongated along the disk are distinguished above and below the disk. A detailed study shows that this is predominantly the substance of the clump. Upon collision with the disk, the stream was divided into three parts: one remained in the same half-space from which the clump originated, the second part pass through the disk, but most of the mass was carried away by the disk's motion. Analysis of the calculation with the clump mass $\ensuremath{M_{c}} = 0.1M_J$ showed that almost the entire clump is absorbed by the disk or remains in its initial half-space, as if ``bouncing'' off the gas disk, which has its own pressure.

The presence of the mentioned hole is also clearly visible in the $xy$ plane, as well as the density waves around it, formed primarily by the clump's material. The disk itself rotates clockwise, and the clump fell toward it and crossed the $xy$ plane near the point (10, 0). Furthermore, some of the material has already spiraled and is falling toward the star. Finally, in the $xz$ slice, the clump itself is clearly visible: it occupies a space larger than one computational cell. A complex gas structure within a radius smaller than $q$ is also visible.

As the clump passes through the disk, two gas flows interact, reducing their velocity. Therefore, accretion toward the star is expected. By the accretion rate, we mean the total flow of matter per unit time passing through the boundary $R = 0.2$~\text{AU}~in the direction toward the star.

\newpage
\def\fracrise{$\dot{M}_{\mathrm{max}}/\dot{M}_{\mathrm{min}}$}
\def\dtrise{$\Delta t_{\mathrm{rise}}$}

\begin{table}[h] \centering
	\caption{Characteristic values and times of change in accretion rates, some of the designations coincide with the table~\ref{tab:models}.
	 \label{tab:accretion} }
\def\McaMa{$M_{ca}^{100}/M_{a}^{100}$}

\begin{tabular}{lccccccc}
	Name  & $\ensuremath{M_{c}}$ &  Nx  & \fracrise & \dtrise & $\ensuremath{M_{c}}^{50}$ & $M_a^{100}$ & \McaMa \\ \hline
	q5i15w90  &    1.0    &  1x  &   418.8   &    3    &      0.93      &    1.57     &   45   \\
	q5i30w90  &    1.0    &  1x  &   237.9   &    4    &      0.90      &    1.32     &   39   \\
	q5i45w90  &    1.0    &  1x  &   107.1   &    4    &      0.85      &    0.99     &   29   \\
	q5i60w90  &    1.0    &  1x  &   57.4    &    5    &      0.79      &    0.70     &   20   \\
	q5i30w90  &    0.3    &  1x  &   234.8   &    4    &      0.28      &    0.58     &   36   \\
	q5i45w90  &    0.3    &  1x  &   109.7   &    4    &      0.28      &     0.5     &   32   \\
	q5i60w90  &    0.3    &  1x  &   52.2    &    5    &      0.27      &    0.38     &   24   \\
	q5i30w90  &    0.1    &  1x  &   42.1    &    4    &     0.096      &    0.148    &   14   \\
	q5i30w15  &    1.0    &  1x  &   42.0    &    4    &      0.88      &    0.33     &   37   \\
	q5i30w45  &    1.0    &  1x  &   89.2    &    6    &      0.94      &    0.51     &   27   \\
	q5i30w135 &    1.0    &  1x  &   39.8    &    4    &      0.78      &    0.69     &   38   \\
	q5i30w165 &    1.0    &  1x  &   23.7    &    6    &      0.75      &    0.38     &   38   \\
	q10i30w90 &    1.0    & 1.5x &   69.2    &    8    &      0.89      &    0.63     &   32   \\
	q15i30w90 &    1.0    &  2x  &   69.5    &   15    &      0.91      &    0.22     &   37   \\
	q5i30w90  &    1.0    &  2x  &   223.1   &    6    &      0.84      &    0.65     &   29   \\
	q10i45w90 &    1.0    & 1.5x &   42.0    &    7    &      0.86      &    0.32     &   26   \\
	q10i45w45 &    1.0    & 1.5x &   39.7    &    5    &      0.88      &    0.16     &   24   \\ \hline
\end{tabular}

{\textbf{Note.}} \fracrise~	is the ratio of the accretion rate at the first maximum to the minimum value, \dtrise~is the time interval in years of the sharp increase in the accretion rate. $\ensuremath{M_{c}}^{50}$~is the mass of the clump captured by the disk or accreted by the time $50t_0$, $M_a^{100}$~is the total mass of the accreted matter (in Jupiter masses) by the time $100t_0$. The last column shows the contribution of the clump matter to $M_a^{100}$ as a percentage. 
\end{table}

%\newpage
Table ~\ref{tab:accretion} presents the behavior of the accretion rate in each of the calculations. We examined the parameter \dtrise~, which is the duration of the accretion rate increase from the minimum value to the first maximum, with the ratio of the maximum to the minimum accretion rate being \fracrise~in this interval. The concept of ``e-folding time'' also appears in the literature for light curves of FU Ori type stars; however, we have not found a clear algorithmic definition of this quantity, so we reserve its use for our calculations for further research.

Furthermore, by analogy with the previous paper~\citep{2024ARep...68..949G}, we estimated the mass of the clump $\ensuremath{M_{c}}^{50}$ remaining in the system at a time 50 years after the start of the calculations, that is, the clump matter moving no faster than the free fall velocity or that has already fallen toward the star through the inner boundary of the calculation region. The last two columns give the total mass $M_a^{100}$ of matter accreted onto the star over 100 years of calculations, as well as the percentage contribution of the clump mass to this value.   
%\newpage
\begin{table}[h] \centering
	\caption{Time moments (in years) from the start of calculations and distances (in AU) from the star at the moments when the center of mass of the clump crosses the plane of the disk.    
		\label{tab:times_distances}
	}
	\begin{tabular}{lcccccccccc}
		Name      & $\ensuremath{M_{c}}$ &  Nx  & $t_1$ & $R_1$ & $t_2$ & $R_2$ & $t_{c1}$ & $R_{c1}$ & $t_{c2}$ & $R_{c2}$ \\ \hline		
		q5i15w90      &    1.0    &  1x  & 5.36  & 10.0  & 12.07 & 10.0  &   5.54   &  10.11   &  12.88   &   6.54   \\
		q5i30w90      &    1.0    &  1x  & 5.36  & 10.0  & 12.07 & 10.0  &   5.61   &   9.91   &  11.93   &   6.32   \\
		q5i45w90      &    1.0    &  1x  & 5.36  & 10.0  & 12.07 & 10.0  &   5.66   &   9.76   &  12.22   &   7.34   \\
		q5i60w90      &    1.0    &  1x  & 5.36  & 10.0  & 12.07 & 10.0  &   5.61   &   9.85   &  12.56   &   8.10   \\
		q5i30w90      &    0.3    &  1x  & 5.36  & 10.0  & 12.07 & 10.0  &   6.06   &   9.25   &  12.92   &   3.80   \\
		q5i45w90      &    0.3    &  1x  & 5.36  & 10.0  & 12.07 & 10.0  &   5.84   &   9.57   &  12.88   &   4.96   \\
		q5i60w90      &    0.3    &  1x  & 5.36  & 10.0  & 12.07 & 10.0  &   5.72   &   9.77   &  13.68   &   6.67   \\
		q5i30w90      &    0.1    &  1x  & 5.36  & 10.0  & 12.07 & 10.0  &    --    &    --    &    --    &    --    \\
		q5i30w15      &    1.0    &  1x  & 8.38  & 5.09  &  395  & 293.5 &   8.69   &   4.64   &    --    &    --    \\
		q5i30w45      &    1.0    &  1x  & 7.61  & 5.86  & 26.6  & 34.1  &   7.89   &   5.56   &    --    &    --    \\
		q5i30w135     &    1.0    &  1x  & 9.82  & 5.86  &  --   &  --   &  10.56   &   5.59   &    --    &    --    \\
		q5i30w165     &    1.0    &  1x  & 9.05  & 5.09  &  --   &  --   &   9.41   &   4.72   &    --    &    --    \\
		q10i30w90     &    1.0    & 1.5x & 7.29  & 20.0  & 26.27 & 20.0  &   7.53   &  20.32   &  26.50   &  14.30   \\
		q15i30w90     &    1.0    &  2x  & 8.82  & 30.0  & 43.69 & 30.0  &   8.87   &  30.01   &  43.53   &  21.27   \\
		q5i30w90      &    1.0    &  2x  & 5.36  & 10.0  & 12.07 & 10.0  &   5.59   &   9.96   &  12.27   &   8.25   \\
		q10i45w90     &    1.0    & 1.5x & 7.29  & 20.0  & 26.27 & 20.0  &   7.84   &  19.87   &  27.02   &  15.66   \\
		q10i45w45     &    1.0    & 1.5x & 13.66 & 11.72 & 67.34 & 68.28 &  14.09   &  11.78   &  94.21   &  53.61   \\ \hline
	\end{tabular}
	
{\textbf{Note.}} Without index~is celestial mechanics approximation, with index ``c''~is gas-dynamic calculations.
\end{table}
%\newpage

Table ~\ref{tab:times_distances} lists the times at which the clump's center of mass crosses the initial disk plane, as well as the distances from the star at these intersections. These data are compared with celestial moments and distances. As we previously found~\citep{2024ARep...68..949G}, the results of gas-dynamic calculations do not agree with the celestial-mechanics approximation due to the presence of gas interactions, which leads to active deceleration of the stream.

\begin{figure}[!ht]\centering
	\includegraphics[width=1.0\linewidth]{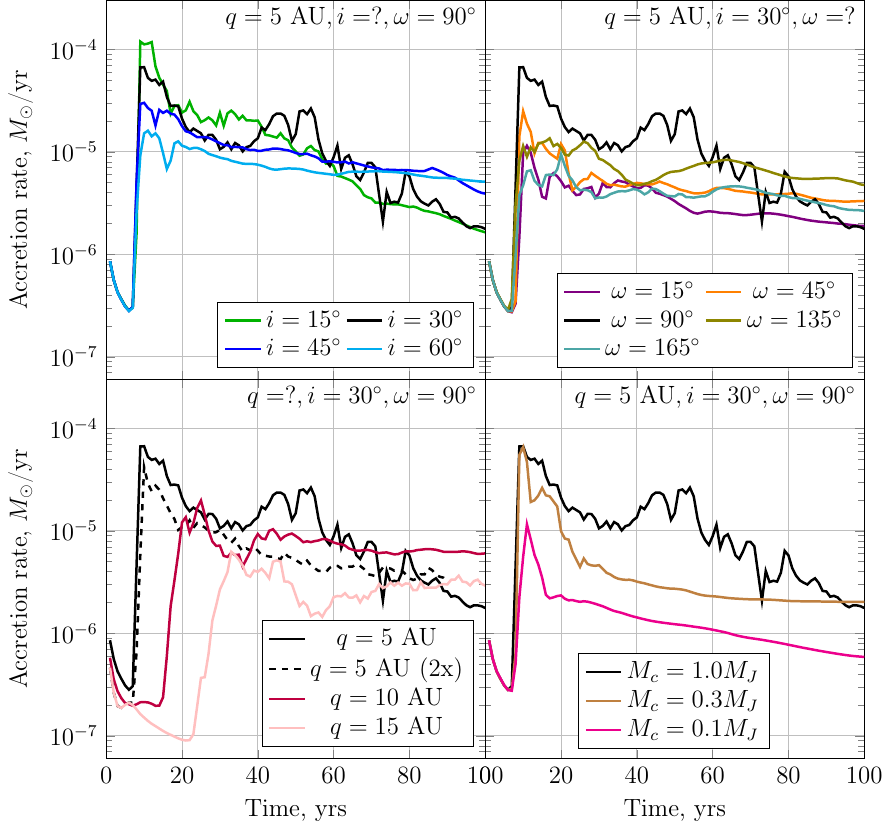}
	\caption{Accretion rate as a function of stream trajectory parameters: top left for varying orbital inclination angle $i$, top right for varying pericenter argument $\omega$, bottom left for varying pericenter distance, bottom right for varying initial clump mass. In all panels, the accretion rate for the reference calculation with parameters $q = 5~\text{AU}, i=30\ensuremath{^\circ}, \omega = 90\ensuremath{^\circ}, \ensuremath{M_{c}} = 1.0 M_J$ is shown in black.}
	\label{fig:accretion} \end{figure}

Fig. ~\ref{fig:accretion} shows accretion rate plots for various parameters. All plots show a common behavior: initial relaxation of gas in the disk, a sharp increase in the accretion rate by several tens of times, a period of maximum accretion (which can be a peak, two-peaked, or plateau-like), and then a gradual decline with
potential oscillations. These oscillations can be caused by either the repeated infall of large remnants of the clump into the disk
(as in the reference calculation, two more accretion peaks are observed) or by smaller condensations already in the disk that stochastically fall toward the star. A detailed analysis of these specific features will be provided below.

\subsection*{Changing the inclination angle of the orbital plane}
The orbital inclination angle $i$ strongly correlates with the accretion rate at maximum brightness (Fig.~\ref{fig:accretion}, top left). This is explained quite simply: the smaller the orbital inclination to the disk plane, the longer the clump's matter interacts with the disk matter and is thus decelerated more strongly. Therefore, the greater part of the gas has a velocity less than the circular velocity, which causes it to fall towards the star.

After the second intersection of the disk plane by the clump's orbit, the clump's matter (and part of the disk's matter) rises above the disk plane in both directions. In the case $i\le 30\ensuremath{^\circ}$, the matter subsequently flows back onto the disk, triggering a repeated burst of accretion. For the case $i = 15\ensuremath{^\circ}$, a repeated increase in the accretion rate is observed at a time of $\sim 20$~years, and for $i=30\ensuremath{^\circ}$ is $\sim 40$~years from the start of the calculations. The total mass of the accreted matter over the calculation period for both models exceeds the initial mass of the clump.
 
The greater the orbital inclination, the more the stream-disk interaction resembles a ``puncture'', meaning the amount of matter undergoing deceleration decreases with increasing $i$. Accordingly, the peak accretion rate is also smaller. For angles $i \ge 45\ensuremath{^\circ}$, accretion is limited to a double (due to two intersections of the disk plane) primary burst.

In numerical terms, this is clearly visible in Table~\ref{tab:accretion}: as the inclination angle increases from
$15\ensuremath{^\circ}$~to~$60\ensuremath{^\circ}$, the \fracrise~ratio drops by a factor of seven, with every $15\ensuremath{^\circ}$~resulting in $\sim 2$ times decrease in this value. This is reliably seen in calculations with fixed $q=5~AU,~ \omega=90\ensuremath{^\circ}$ for two clump masses ($1M_J$ and $0.3M_J$), as well as for $q=10~AU,~\omega=90\ensuremath{^\circ}$.

The fraction of the clump's matter remaining in the system and the mass of the accreted matter (the percentage of the clump's matter in it) also decrease significantly with increasing orbital inclination.

\subsection*{Variation of the argument of periapsis}

Changing the argument of periapsis $\omega$ has a less significant effect on the accretion rate than the orbital inclination, however,
the following features can be noticed (see Fig.~\ref{fig:accretion} top right).

First, calculations with $\omega = 135\ensuremath{^\circ}$ and $\omega = 165\ensuremath{^\circ}$ yield only one intersection of the disk plane, which is clearly seen in Table~\ref{tab:times_distances}. Due to the formulation of maximally homogeneous initial conditions, the second intersection should have occurred ``in the past''. For $\omega = 135\ensuremath{^\circ}$, it should have occurred at a distance of $\sim 35$~\text{AU}, which would most likely have caused an accretion burst with a peak of less than $5 \times 10^{-6} M_{\odot}$/yr (see below). For $\omega = 165\ensuremath{^\circ}$, the first intersection should have occurred at a distance of about $300$~\text{AU}, where, according to our numerical model, the disk as such no longer exists. Both of these considerations allow us to neglect the influence of failed intersections on the behavior of the accretion rate.

For $\omega = 15\ensuremath{^\circ}$ and $\omega = 165\ensuremath{^\circ}$, the situation is geometrically symmetric: in both cases, the orbital pericenter and the disk plane are separated by $15\ensuremath{^\circ}$, so the nodes of the initial orbit are asymmetric with respect to the star (one is significantly closer, and the other is significantly further). A closer intersection produces a brighter accretion burst, but since a closer intersection occurs later for a larger argument of pericenter, the accretion peak for $\omega = 165\ensuremath{^\circ}$ is observed later than for $\omega = 15\ensuremath{^\circ}$. For $\omega = 45\ensuremath{^\circ}$ and $\omega = 135\ensuremath{^\circ}$ the situation is similar, but due to the higher relative collision velocity, the accretion rate is higher at the peak than for $\omega = 15\ensuremath{^\circ}$. For $\omega = 135\ensuremath{^\circ}$, a single intersection of the disk plane causes significant deceleration of the stream matter; by this point, the stream itself becomes more elongated than for $\omega = 45\ensuremath{^\circ}$, which leads to the appearance of a ``plateau'' in the accretion rate dependence on time.
At $\omega = 90\ensuremath{^\circ}$, both peaks from the two crossings merge with each other and are indistinguishable, since both crossings occur within a short period of time (see table~\ref{tab:times_distances}), and more efficient deceleration increases the mass moving toward the star.

The largest amount of clump matter remains in the system at $\omega = 45\ensuremath{^\circ}$ (see table~\ref{tab:accretion}),
while the percentage of clump matter accreted onto the star is smallest in this case. This is evident from calculations
with $q = 5~\text{AU}$ and with $q = 10~\text{AU}$. With this pericenter argument, the most effective deceleration of the clump
to the disk is achieved: the clump is relatively close to the star when it crosses the disk plane (closer than for $\omega = 90\ensuremath{^\circ}$, see Table~\ref{tab:times_distances}) and is quite compact, since  the intersection time
is less than for $\omega = 15\ensuremath{^\circ}$, and much less than for $\omega > 90\ensuremath{^\circ}$.

The mass of the accreted matter decreases significantly as $\omega$ deviates from $90\ensuremath{^\circ}$. Obviously, the closer to the star the two encounters between the initial orbit of the clump and the disk plane occur, the more matter can fall onto the star. At the same time, the percentage contribution of the clump's matter to this mass depends weakly on $\omega$.

\subsection*{Change in pericentric distance}

Variations in the orbital pericentric distance lead to noticeable differences in the maximum accretion rates (Fig.~\ref{fig:accretion}, lower left). The gas stream, interacting with a more distant and rarefied region of the disk, produces an accretion burst later than during a closer flyby. The change in accretion rate is generally comparable: in any case, the difference from the minimum to the peak is at least 60 times (see Table~\ref{tab:accretion}). The lower accretion rate before the burst in higher-resolution calculations is explained by the fact that regions close to the inner boundary along $R$ are better resolved in the vertical direction (the influence of numerical effects is less pronounced), while the density in this direction initially decreases according to the barometric law. In addition, the distant fall causes a smoother rise in the accretion rate to the peak, because the resulting spiral wave becomes more elongated, longer and smoother ``flows'' towards the star, which increases the plateau at the peak in the accretion rate plot.

These plots allow us to say that the maximum accretion rate decreases by approximately half an order of magnitude with a twice increase in the pericentric distance and by an order of magnitude with a three-fold increase.

It's interesting to note that the reference calculation, performed at ``double'' resolution, gave a somewhat different accretion pattern.
This is because the clump also occupied one cell, but the cell itself was half the size, yielding an eight times higher density. Because of this, the stream is thinner and more compact, meaning its matter is decelerated less by the disk compared to a less dense clump of the same mass. Therefore, a second accretion peak is not observed in the plot. However, due to the identical orbit, the accretion burst \fracrise~is practically the same as in the reference calculation.

Note that the calculation with $q = 15$~\text{AU}~gives the first intersection of the disk plane at a distance of $30$~\text{AU}~(see table~\ref{tab:times_distances}), which is closer than the missed intersection at $\omega = 135\ensuremath{^\circ}$ (see above). The accretion rate at the peak does not exceed $5 \times 10^{-6} M_{\odot}$/yr, meaning that the failed intersection would have an even lower accretion rate.

The pericentric distance, all other things being equal, has no effect on how much of the clump's matter remains in the system: it is primarily determined by where and when the disk plane is crossed. At the same time, the mass of accreted matter decreases significantly with increasing pericentric distance over the calculation time. On the one hand, this is apparently due to the overall decrease in the accretion rate with increasing $q$. On the other hand, the development times of the disk's gas-dynamic response to a more distant collision are longer. The percentage contribution of clump matter to the total accreted mass is independent of $q$.

\subsection*{Variation in the mass of the falling clump}

Finally, a decrease in the clump mass predictably leads to a lower accretion rate (Fig.~\ref{fig:accretion}, bottom right),
and more efficient capture of clump matter (the parameter $\ensuremath{M_{c}}^{50}$ is at least 90\% of the clump mass, all other things being equal). At $\ensuremath{M_{c}} = 0.3M_J$, the maximum accretion rate reaches the same value as for a single Jupiter mass, but the peak itself is narrower, the accretion rate declines more rapidly, and the disk manages to return to a quiescent state more quickly, so no fluctuations in the accretion rate are observed after the outburst. The same is true for orbits with large inclination angles: in Table~\ref{tab:accretion}, the \fracrise~parameter for identical angles, regardless of the clump mass, is very close.
A similar picture is observed when a gas stream 10 times less massive than Jupiter. Interestingly, there is no intersection of the disk plane (see Table~\ref{tab:times_distances}). This resembles the clump ``bouncing off'' the disk, like a pebble bouncing off water. However, in our case, almost the entire clump ($96\%$ of its initial mass) enters the disk from only one side and is carried away by its rotation. Moreover, in this model, the contribution of the clump's matter to the accreted mass is minimal.

For models with $\ensuremath{M_{c}} \le 0.3$, the mass of the accreted matter exceeds the initial mass of the clump by 1.5--2 times. The contribution of the clump matter to $M_a^{100}$ decreases with decreasing initial mass.

\section*{DISCUSSION}
\noindent

It is interesting to compare the accretion rates with the best-known light curves of FU Ori type stars. It is assumed that the brightness variations are due to the accretion luminosity, which, other parameters being constant, is directly proportional to the accretion rate. The contribution of the disk to the emission in the V and B optical bands under consideration is assumed to be negligible. The response of the circumstellar disk to a FU~Ori like outburst was considered in~\citet{2025A&A...700L..24L}. It was shown that optically thin regions of the disk heat and cool almost instantaneously. Heating of optically thick layers of the disk requires several years, and cooling takes about a decade. However, these regions make a significant contribution to the emission in the near- and mid-IR ranges. It should be noted that the authors did not take into account the change in the vertical density distribution with disk temperature variations, so the relaxation times of optically thick layers of the disk may be shorter due to this effect. Thus, in~\citet{2017A&A...605A..30M} it was shown that the characteristic time of thermal relaxation of the matter of a disk with a mass of $0.01 M_\odot$ under the influence of radiative processes at a distance of $>3$ AU is less than a year, and this value decreases with increasing distance and temperature. We also note that for a characteristic density of $10^{-10}$ g~cm$^{-3}$, the circumstellar disk is optically thin at temperatures in the range of $900-8000$~K, with an increase in density to values of $10^{-6}$ g~cm$^{-3}$, the temperature range narrows to $1500-4000$~K~\citep{2007A&A...475...37S}. Thus, the inner regions of the disk in the considered model can be considered optically thin.

\begin{figure}[!ht]\centering
	\includegraphics[width=1.0\linewidth, page=3]{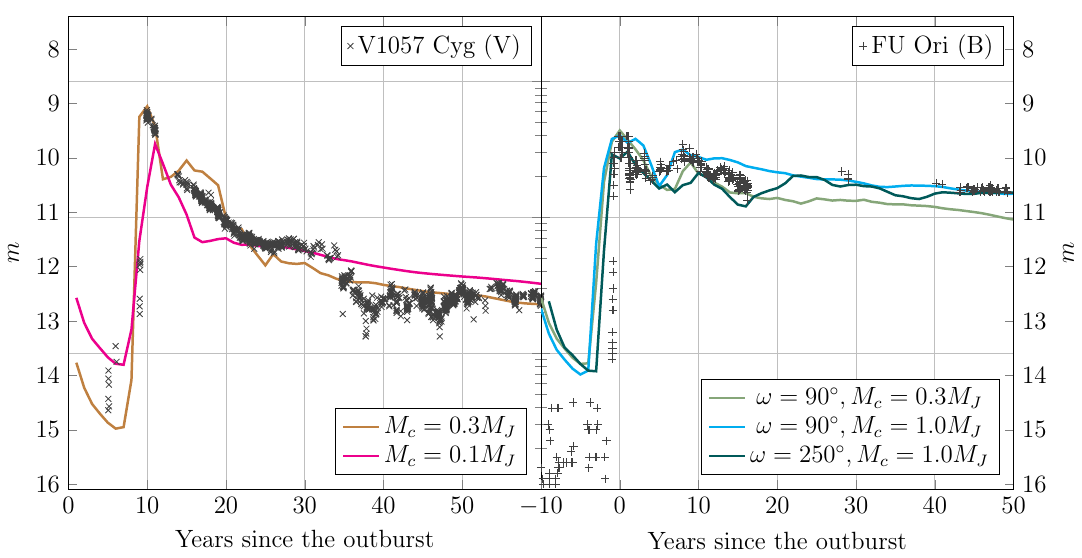}
	\caption{Accretion rate in calculated models compared with light curves for the star V1057~Cyg (in the V band) and for the star FU~Ori (in the B band). The data are taken from~\citet{1954ZA.....35...74W,1968ApJ...151..977M,1983SvAL....9...92K,2006ApJ...648.1099G,2021ApJ...917...80S}. For the first star, the best fit was obtained with parameters
$q = 5~\text{AU}, i=30\ensuremath{^\circ}, \omega=90\ensuremath{^\circ}$; the mass is given in the legend. For the second star~is $q=5~\text{AU}, i=60\ensuremath{^\circ}$,
the other parameters are given in the legend. The calculation with $q=5~\text{AU}, i=60\ensuremath{^\circ}, \omega =250\ensuremath{^\circ}$~is the DP-r60 calculation from the previous paper~\citep{2024ARep...68..949G}. Some accretion rate graphs have been shifted along the axes without scaling for better clarity.}
	\label{fig:V1057CygFUOri} 
\end{figure}

Figure~\ref{fig:V1057CygFUOri} (left) shows the V-band light curve of V1057~Cyg, constructed using the data from the Appendix to~\citet{2021ApJ...917...80S}. The accretion rate plots, which, in our opinion, are most similar to it, are superimposed on this light curve. Calculations with the parameters $q=5$~\text{AU}, $i=30\ensuremath{^\circ}$, $\omega=90\ensuremath{^\circ}$, $\ensuremath{M_{c}}=0.3M_J$ and $\ensuremath{M_{c}}=0.1M_J$ are shown. The maximum accretion rate in the first calculation reaches $6.6 \times 10^{-5} M_\odot$/year, in the second is $1.2 \times 10^{-5} M_\odot$/year.

Figure~\ref{fig:V1057CygFUOri} (right) shows the B-band light curve of the FU~Ori star; the data for it were taken from~\citet{1954ZA.....35...74W,1968ApJ...151..977M,1983SvAL....9...92K,2006ApJ...648.1099G}. Also superimposed on it are plots of the accretion rate, which, in our opinion, are most similar to the FU~Ori light curve. Shown are two calculations from the current paper, differing in the initial mass of the clump, with the same orbital parameters $q=5~\text{AU}, i=60\ensuremath{^\circ}, \omega=90\ensuremath{^\circ}$, as well as a calculation from the previous paper (with $\omega = 250\ensuremath{^\circ}$). For the case $\ensuremath{M_{c}} = 0.3M_J$ the maximum accretion rate has a value of $1.5\times 10^{-5}M_\odot$/year, in the case $\ensuremath{M_{c}} = 1M_J$ --- $1.6\times 10^{-5}$/year.

Accretion rates for these two objects were estimated by different authors based on modeling the spectral energy distribution and comparing it with observations. Key parameters in these models include the inclination of the accretion disk to the plane of the sky and the mass of the star. Because these parameters are not precisely determined, there is a wide range of values for the maximum accretion rates of the objects under consideration.

The mass of the star V1057~Cyg is estimated to be $\sim 0.5 M_\odot$~\citep{2006ApJ...648.1099G}, the disk inclination angle in the literature
varies from $0\ensuremath{^\circ}$ to $62\ensuremath{^\circ}$. This gives an estimate of the maximum accretion rate in the range from $4\times 10^{-5}$~\citep{2004A&A...422..171L} to $2\times 10^{-3} M_\odot$/yr~\citep{2021ApJ...917...80S}. In a number of studies, the maximum accretion rate is close to $\sim 10^{-4} M_\odot$/yr~\citep{1988ApJ...325..231K,1996ApJ...473..422P,2014AJ....147..140G}.

In the case of FU Ori, the star's mass was estimated by various authors to be in the range of $0.3-1 M_\odot$~\citep{1988ApJ...325..231K,1996ApJ...473..422P,2001A&A...375..455L,2006ApJ...648.1099G}, while the disk inclination was determined to be in the range of
$50\ensuremath{^\circ}-60\ensuremath{^\circ}$. The accretion rate at the maximum was determined to be in the range of $10^{-5}$~\citep{2014AJ....147..140G}  to $3\times 10^{-4} M_{\odot}$/yr~\citep{1996ApJ...473..422P,2006ApJ...648.1099G}. An estimate of the order of $\sim 10^{-5} M_\odot$/year was also obtained in~\citet{2020ApJ...889...59P,2022A&A...663A..86L}.

Thus, the maximum accretion rate values obtained in our calculations ($6.6 \times 10^{-5} M_{\odot}$/yr for V1057~Cyg and $1.5 \times 10^{-5} M_\odot$/yr for FU~Ori) are within the range of estimates presented in the literature. It is interesting to note that the different behavior of the accretion rate after the outburst for the two objects under consideration can be described within our model by variations in the inclination angle of the initial stream orbit. The smaller the initial inclination angle of the stream orbit, the more rapidly the accretion rate decreases after the outburst. Thus, in the case of V1057~Cyg, the shape of the light curve shows a sharp spike followed by a rapid decline in luminosity, which is consistent with the behavior of the accretion rate for models with the parameter $i=30\ensuremath{^\circ}$. In the case of FU~Ori, the luminosity after the outburst decreases more slowly, according to calculations with $i=60\ensuremath{^\circ}$. Furthermore, immediately after the outburst, fluctuations are noticeable, which are present in the accretion rate plots. According to our calculations, the accretion rate at maximum increases with decreasing initial orbital inclination, which is also consistent with the fact that the literature estimate of the accretion rate at maximum brightness for V1057~Cyg significantly exceeds that for FU~Ori.

Apparently, variations in the mass of the infalling clump within the range of $0.3-1M_J$ affect only the duration of the outburst, but not its amplitude, which is primarily determined by the orbital inclination. Furthermore, not just one clump but several may fall onto the protoplanetary disk, either moving along the same trajectory with some time delay or along different trajectories. This circumstance may explain the various plateaus and secondary dips in the light curves of FU~Ori type stars.

\section*{CONCLUSION}
The paper continues a series of studies simulation the accretion of matter from the external environment onto the inner regions of a circumstellar disk. It was shown that the parameters of the initial stream orbit significantly affect the magnitude of the accretion rate burst, as well as the shape of its time dependence. The accretion rate at the moment of the burst increases with decreasing orbital inclination and pericentric distance, as well as with increasing initial stream mass. An increase in the accretion rate by two orders of magnitude was obtained for models with parameters for which the orbital inclination angle $i<45^\circ$, pericentric distance $q<10$~AU, and mass $M_c>0.1 M_J$. The time it takes for the accretion rate to reach its maximum value increases with the distance from the stream-disk collision region (or pericentric distance) from 3 to 15 years. The nature of the accretion rate decline after the outburst depends significantly on the initial inclination of the stream's orbit, as well as its initial mass: the smaller the inclination and mass, the faster the accretion rate declines over time. It should also be noted that the fraction of disk matter accreted onto the star over the simulation period exceeds the contribution of the clump's material for all models. 

\section*{ACKNOWLEDGMENTS}
The calculations were carried out using the resources of
the Joint SuperComputer Center of the Russian Academy
of Sciences Branch of Federal State Institution ``Scientific
Research Institute for System Analysis of the Russian Academy of Sciences'' (Savin et al., 2019). 

\bibliographystyle{rusnat}
\bibliography{GD24}

@ARTICLE{2009ApJ...700.1502A,
       author = {{Andrews}, Sean M. and {Wilner}, D.~J. and {Hughes}, A.~M. and {Qi}, Chunhua and {Dullemond}, C.~P.},
        title = "{Protoplanetary Disk Structures in Ophiuchus}",
      journal = {\apj},
         year = 2009,
        month = aug,
       volume = {700},
       number = {2},
        pages = {1502-1523},
          doi = {10.1088/0004-637X/700/2/1502},
archivePrefix = {arXiv},
       eprint = {0906.0730},
 primaryClass = {astro-ph.EP},
       adsurl = {https://ui.adsabs.harvard.edu/abs/2009ApJ...700.1502A}
}

@ARTICLE{ShakuraSunyaev1973,
	author = {{Shakura}, N.~I. and {Sunyaev}, R.~A.},
	title = "{Black holes in binary systems. Observational appearance.}",
	journal = {Astron. Astrophys.}, 
	year = 1973,
	month = jan,
	volume = {24},
	pages = {337-355},
	adsurl = {https://ui.adsabs.harvard.edu/abs/1973A&A....24..337S}
}

@ARTICLE{2018MNRAS.475.2642K,
       author = {{Kuffmeier}, Michael and {Frimann}, S{\o}ren and {Jensen}, Sigurd S. and {Haugb{\o}lle}, Troels},
        title = "{Episodic accretion: the interplay of infall and disc instabilities}",
      journal = {\mnras},
         year = 2018,
        month = apr,
       volume = {475},
       number = {2},
        pages = {2642-2658},
          doi = {10.1093/mnras/sty024},
archivePrefix = {arXiv},
       eprint = {1710.00931},
 primaryClass = {astro-ph.SR},
       adsurl = {https://ui.adsabs.harvard.edu/abs/2018MNRAS.475.2642K}
}

@ARTICLE{1997ApJ...486..372B,
       author = {{Bell}, K.~R. and {Cassen}, P.~M. and {Klahr}, H.~H. and {Henning}, Th.},
        title = "{The Structure and Appearance of Protostellar Accretion Disks: Limits on Disk Flaring}",
      journal = {\apj},
         year = 1997,
        month = sep,
       volume = {486},
       number = {1},
        pages = {372-387},
          doi = {10.1086/304514},
       adsurl = {https://ui.adsabs.harvard.edu/abs/1997ApJ...486..372B}
}

@ARTICLE{1997ApJ...490..368C,
       author = {{Chiang}, E.~I. and {Goldreich}, P.},
        title = "{Spectral Energy Distributions of T Tauri Stars with Passive Circumstellar Disks}",
      journal = {\apj},
         year = 1997,
        month = nov,
       volume = {490},
       number = {1},
        pages = {368-376},
          doi = {10.1086/304869},
archivePrefix = {arXiv},
       eprint = {astro-ph/9706042},
 primaryClass = {astro-ph},
       adsurl = {https://ui.adsabs.harvard.edu/abs/1997ApJ...490..368C}
}

@ARTICLE{2023PCT...1868..269G,
       author = {{Grigoryev}, V. and {Demidova}, T.},
        title = "{Comparison of two methods for modelling the dynamics
of gas flows in a protoplanetary disk}",
      journal = {Commun. Comput. In-
form. Sci.},
         year = 2023,
        month = nov,
       volume = {1868},
       number = {},
        pages = {269},
          doi = {10.1007/978-3-031-38864-4_19},
}

@ARTICLE{2024ARep...68..949G,
       author = {{Grigoryev}, V.~V. and {Demidova}, T.~V.},
        title = "{Simulation of the Free Fall of a Gas Stream on a Protoplanetary Disk}",
      journal = {Astronomy Reports},
         year = 2024,
        month = oct,
       volume = {68},
       number = {10},
        pages = {949-966},
          doi = {10.1134/S1063772924700859},
archivePrefix = {arXiv},
       eprint = {2503.20606},
 primaryClass = {astro-ph.SR},
       adsurl = {https://ui.adsabs.harvard.edu/abs/2024ARep...68..949G}
}

@ARTICLE{2024Proc...40G,
       author = {{Grigoryev}, V. and {Demidova}, T.},
        title = "{Simulation of stream falling onto a protoplanetary disk}",
      journal = {in Proceedings
of the 11th International Youth School-Seminar
of E. V. Voskresenskii on Mathematical Model-
ing, Numerical Methods and Software Packages,
Saransk},
         year = 2024,
        month = jul,
       volume = {},
       number = {},
        pages = {40},
          doi = {10.1007/978-3-031-38864-4_19},
}

@ARTICLE{2007ApJS..170..228M,
       author = {{Mignone}, A. and {Bodo}, G. and {Massaglia}, S. and {Matsakos}, T. and {Tesileanu}, O. and {Zanni}, C. and {Ferrari}, A.},
        title = "{PLUTO: A Numerical Code for Computational Astrophysics}",
      journal = {\apjs},
         year = 2007,
        month = may,
       volume = {170},
       number = {1},
        pages = {228-242},
          doi = {10.1086/513316},
archivePrefix = {arXiv},
       eprint = {astro-ph/0701854},
 primaryClass = {astro-ph},
       adsurl = {https://ui.adsabs.harvard.edu/abs/2007ApJS..170..228M}
}

@ARTICLE{Braginskii1965en,
	author = {{Braginskii}, S.~I.},
	title = "{Transport Processes in a Plasma}",
	journal = {Reviews of Plasma Physics},
	year = 1965,
	month = jan,
	volume = {1},
	pages = {205},
	adsurl = {https://ui.adsabs.harvard.edu/abs/1965RvPP....1..205B}
}

@ARTICLE{2024AstL...50..625D,
       author = {{Demidova}, T.~V. and {Grigoryev}, V.~V.},
        title = "{Simulation of Images of Protoplanetary Disks after Collision with Free-Floating Planets}",
      journal = {Astronomy Letters},
         year = 2024,
        month = oct,
       volume = {50},
       number = {10},
        pages = {625-637},
          doi = {10.1134/S1063773724700476},
archivePrefix = {arXiv},
       eprint = {2506.23795},
 primaryClass = {astro-ph.EP},
       adsurl = {https://ui.adsabs.harvard.edu/abs/2024AstL...50..625D}
}

@ARTICLE{2023ApJ...953...38D,
       author = {{Demidova}, Tatiana V. and {Grinin}, Vladimir P.},
        title = "{Three-dimensional SPH Simulations of FU Orionis Star Flares in the Clumpy Accretion Model}",
      journal = {\apj},
         year = 2023,
        month = aug,
       volume = {953},
       number = {1},
          eid = {38},
        pages = {38},
          doi = {10.3847/1538-4357/acdf5f},
archivePrefix = {arXiv},
       eprint = {2308.04936},
 primaryClass = {astro-ph.SR},
       adsurl = {https://ui.adsabs.harvard.edu/abs/2023ApJ...953...38D}
}

@ARTICLE{1994A&A...286..149D,
       author = {{Dutrey}, A. and {Guilloteau}, S. and {Simon}, M.},
        title = "{Images of the GG Tauri rotating ring}",
      journal = {\aap},
         year = 1994,
        month = jun,
       volume = {286},
        pages = {149-159},
       adsurl = {https://ui.adsabs.harvard.edu/abs/1994A&A...286..149D}
}

@ARTICLE{2023ApJ...958...98F,
       author = {{Flores}, Christian and {Ohashi}, Nagayoshi and {Tobin}, John J. and {J{\o}rgensen}, Jes K. and {Takakuwa}, Shigehisa and {Li}, Zhi-Yun and {Lin}, Zhe-Yu Daniel and {van't Hoff}, Merel L.~R. and {Plunkett}, Adele L. and {Yamato}, Yoshihide and {Sai (Insa Choi)}, Jinshi and {Koch}, Patrick M. and {Yen}, Hsi-Wei and {Aikawa}, Yuri and {Aso}, Yusuke and {de Gregorio-Monsalvo}, Itziar and {Kido}, Miyu and {Kwon}, Woojin and {Lee}, Jeong-Eun and {Lee}, Chang Won and {Looney}, Leslie W. and {Santamar{\'\i}a-Miranda}, Alejandro and {Sharma}, Rajeeb and {Thieme}, Travis J. and {Williams}, Jonathan P. and {Han}, Ilseung and {Narayanan}, Suchitra and {Lai}, Shih-Ping},
        title = "{Early Planet Formation in Embedded Disks (eDisk). XII. Accretion Streamers, Protoplanetary Disk, and Outflow in the Class I Source Oph IRS 63}",
      journal = {\apj},
         year = 2023,
        month = nov,
       volume = {958},
       number = {1},
          eid = {98},
        pages = {98},
          doi = {10.3847/1538-4357/acf7c1},
archivePrefix = {arXiv},
       eprint = {2310.14617},
 primaryClass = {astro-ph.SR},
       adsurl = {https://ui.adsabs.harvard.edu/abs/2023ApJ...958...98F}
}

@ARTICLE{2006ApJ...648.1099G,
       author = {{Green}, J.~D. and {Hartmann}, L. and {Calvet}, N. and {Watson}, D.~M. and {Ibrahimov}, M. and {Furlan}, E. and {Sargent}, B. and {Forrest}, W.~J.},
        title = "{Spitzer IRS Observations of FU Orionis Objects}",
      journal = {\apj},
         year = 2006,
        month = sep,
       volume = {648},
       number = {2},
        pages = {1099-1109},
          doi = {10.1086/505932},
archivePrefix = {arXiv},
       eprint = {astro-ph/0605365},
 primaryClass = {astro-ph},
       adsurl = {https://ui.adsabs.harvard.edu/abs/2006ApJ...648.1099G}
}

@ARTICLE{2024ApJ...966...96H,
       author = {{Hales}, A.~S. and {Gupta}, A. and {Ru{\'\i}z-Rodr{\'\i}guez}, D. and {Williams}, J.~P. and {P{\'e}rez}, S. and {Cieza}, L. and {Gonz{\'a}lez-Ruilova}, C. and {Pineda}, J.~E. and {Santamar{\'\i}a-Miranda}, A. and {Tobin}, J. and {Weber}, P. and {Zhu}, Z. and {Zurlo}, A.},
        title = "{Discovery of an Accretion Streamer and a Slow Wide-angle Outflow around FU Orionis}",
      journal = {\apj},
         year = 2024,
        month = may,
       volume = {966},
       number = {1},
          eid = {96},
        pages = {96},
          doi = {10.3847/1538-4357/ad31a1},
archivePrefix = {arXiv},
       eprint = {2405.03033},
 primaryClass = {astro-ph.SR},
       adsurl = {https://ui.adsabs.harvard.edu/abs/2024ApJ...966...96H}
}

@ARTICLE{1985ApJ...299..462H,
       author = {{Hartmann}, L. and {Kenyon}, S.~J.},
        title = "{On the nature of FU Orionis objects.}",
      journal = {\apj},
         year = 1985,
        month = dec,
       volume = {299},
        pages = {462-478},
          doi = {10.1086/163713},
       adsurl = {https://ui.adsabs.harvard.edu/abs/1985ApJ...299..462H}
}

@ARTICLE{1996ARA&A..34..207H,
       author = {{Hartmann}, Lee and {Kenyon}, Scott J.},
        title = "{The FU Orionis Phenomenon}",
      journal = {\araa},
         year = 1996,
        month = jan,
       volume = {34},
        pages = {207-240},
          doi = {10.1146/annurev.astro.34.1.207},
       adsurl = {https://ui.adsabs.harvard.edu/abs/1996ARA&A..34..207H}
}

@ARTICLE{2024A&A...686A.246H,
       author = {{Heigl}, S. and {Hoemann}, E. and {Burkert}, A.},
        title = "{Protostellar disk accretion in turbulent filaments}",
      journal = {\aap},
         year = 2024,
        month = jun,
       volume = {686},
          eid = {A246},
        pages = {A246},
          doi = {10.1051/0004-6361/202449154},
archivePrefix = {arXiv},
       eprint = {2401.03779},
 primaryClass = {astro-ph.GA},
       adsurl = {https://ui.adsabs.harvard.edu/abs/2024A&A...686A.246H}
}

@ARTICLE{1987ApJ...323..714K,
       author = {{Kenyon}, S.~J. and {Hartmann}, L.},
        title = "{Spectral Energy Distributions of T Tauri Stars: Disk Flaring and Limits on Accretion}",
      journal = {\apj},
         year = 1987,
        month = dec,
       volume = {323},
        pages = {714},
          doi = {10.1086/165866},
       adsurl = {https://ui.adsabs.harvard.edu/abs/1987ApJ...323..714K}
}

@ARTICLE{1988ApJ...325..231K,
       author = {{Kenyon}, S.~J. and {Hartmann}, L. and {Hewett}, R.},
        title = "{Accretion Disk Models for FU Orionis and V1057 Cygni: Detailed Comparisons between Observations and Theory}",
      journal = {\apj},
         year = 1988,
        month = feb,
       volume = {325},
        pages = {231},
          doi = {10.1086/165999},
       adsurl = {https://ui.adsabs.harvard.edu/abs/1988ApJ...325..231K}
}

@ARTICLE{1983SvAL....9...92K,
       author = {{Kolotilov}, E.~A. and {Petrov}, P.~P.},
        title = "{Studies of the Fu-Orionis Stars - Part One - is V1515-CYGNI Starting to Fade}",
      journal = {Soviet Astronomy Letters},
         year = 1983,
        month = feb,
       volume = {9},
        pages = {92-94},
       adsurl = {https://ui.adsabs.harvard.edu/abs/1983SvAL....9...92K}
}

@ARTICLE{2004A&A...422..171L,
       author = {{Lachaume}, R.},
        title = "{The vertical structure of T Tauri accretion discs. IV. Self-irradiation of the disc in the FU Orionis outburst phase}",
      journal = {\aap},
         year = 2004,
        month = jul,
       volume = {422},
        pages = {171-176},
          doi = {10.1051/0004-6361:20040287},
archivePrefix = {arXiv},
       eprint = {astro-ph/0404094},
 primaryClass = {astro-ph},
       adsurl = {https://ui.adsabs.harvard.edu/abs/2004A&A...422..171L}
}

@ARTICLE{2025A&A...700L..24L,
       author = {{Laznevoi}, S.~I. and {Akimkin}, V.~V. and {Pavlyuchenkov}, Ya. N. and {Il'in}, V.~B. and {K{\'o}sp{\'a}l}, {\'A}. and {{\'A}brah{\'a}m}, P.},
        title = "{Time-dependent response of protoplanetary disk temperature to an FU Ori-type luminosity outburst}",
      journal = {\aap},
         year = 2025,
        month = aug,
       volume = {700},
          eid = {L24},
        pages = {L24},
          doi = {10.1051/0004-6361/202554962},
archivePrefix = {arXiv},
       eprint = {2508.04686},
 primaryClass = {astro-ph.EP},
       adsurl = {https://ui.adsabs.harvard.edu/abs/2025A&A...700L..24L}
}

@ARTICLE{2001A&A...375..455L,
       author = {{Lodato}, G. and {Bertin}, G.},
        title = "{The spectral energy distribution of self-gravitating protostellar disks}",
      journal = {\aap},
         year = 2001,
        month = aug,
       volume = {375},
        pages = {455-468},
          doi = {10.1051/0004-6361:20010868},
archivePrefix = {arXiv},
       eprint = {astro-ph/0106325},
 primaryClass = {astro-ph},
       adsurl = {https://ui.adsabs.harvard.edu/abs/2001A&A...375..455L}
}

@ARTICLE{2022A&A...663A..86L,
       author = {{Lykou}, F. and {{\'A}brah{\'a}m}, P. and {Chen}, L. and {Varga}, J. and {K{\'o}sp{\'a}l}, {\'A}. and {Matter}, A. and {Siwak}, M. and {Szab{\'o}}, Zs. M. and {Zhu}, Z. and {Liu}, H.~B. and {Lopez}, B. and {Allouche}, F. and {Augereau}, J.-C. and {Berio}, P. and {Cruzal{\`e}bes}, P. and {Dominik}, C. and {Henning}, Th. and {Hofmann}, K.-H. and {Hogerheijde}, M. and {Jaffe}, W.~J. and {Kokoulina}, E. and {Lagarde}, S. and {Meilland}, A. and {Millour}, F. and {Pantin}, E. and {Petrov}, R. and {Robbe-Dubois}, S. and {Schertl}, D. and {Scheuck}, M. and {van Boekel}, R. and {Waters}, L.~B.~F.~M. and {Weigelt}, G. and {Wolf}, S.},
        title = "{The disk of FU Orionis viewed with MATISSE/VLTI. First interferometric observations in L and M bands}",
      journal = {\aap},
         year = 2022,
        month = jul,
       volume = {663},
          eid = {A86},
        pages = {A86},
          doi = {10.1051/0004-6361/202142788},
archivePrefix = {arXiv},
       eprint = {2205.10173},
 primaryClass = {astro-ph.SR},
       adsurl = {https://ui.adsabs.harvard.edu/abs/2022A&A...663A..86L}
}

@ARTICLE{2017A&A...605A..30M,
       author = {{Malygin}, M.~G. and {Klahr}, H. and {Semenov}, D. and {Henning}, Th. and {Dullemond}, C.~P.},
        title = "{Efficiency of thermal relaxation by radiative processes in protoplanetary discs: constraints on hydrodynamic turbulence}",
      journal = {\aap},
         year = 2017,
        month = sep,
       volume = {605},
          eid = {A30},
        pages = {A30},
          doi = {10.1051/0004-6361/201629933},
archivePrefix = {arXiv},
       eprint = {1704.06786},
 primaryClass = {astro-ph.EP},
       adsurl = {https://ui.adsabs.harvard.edu/abs/2017A&A...605A..30M}
}

@ARTICLE{2025MNRAS.543.3321M,
       author = {{Mayer}, Alexander C. and {Naab}, Thorsten and {Caselli}, Paola and {Ivlev}, Alexei V. and {Grassi}, Tommaso and {Zier}, Oliver and {Pakmor}, Rudiger and {Walch}, Stefanie and {Springel}, Volker},
        title = "{Protostellar discs in their natural habitat ? the formation of protostars and their accretion discs in the turbulent and magnetized interstellar medium}",
      journal = {\mnras},
         year = 2025,
        month = nov,
       volume = {543},
       number = {4},
        pages = {3321-3344},
          doi = {10.1093/mnras/staf1404},
archivePrefix = {arXiv},
       eprint = {2506.14394},
 primaryClass = {astro-ph.GA},
       adsurl = {https://ui.adsabs.harvard.edu/abs/2025MNRAS.543.3321M}
}

@ARTICLE{1968ApJ...151..977M,
       author = {{Mendoza V.}, Eugenio E.},
        title = "{Infrared Excesses in T Tauri Stars and Related Objects}",
      journal = {\apj},
         year = 1968,
        month = mar,
       volume = {151},
        pages = {977},
          doi = {10.1086/149497},
       adsurl = {https://ui.adsabs.harvard.edu/abs/1968ApJ...151..977M}
}

@ARTICLE{2020ApJ...889...59P,
       author = {{P{\'e}rez}, Sebasti{\'a}n and {Hales}, Antonio and {Liu}, Hauyu Baobab and {Zhu}, Zhaohuan and {Casassus}, Simon and {Williams}, Jonathan and {Zurlo}, Alice and {Cuello}, Nicol{\'a}s and {Cieza}, Lucas and {Principe}, David},
        title = "{Resolving the FU Orionis System with ALMA: Interacting Twin Disks?}",
      journal = {\apj},
         year = 2020,
        month = jan,
       volume = {889},
       number = {1},
          eid = {59},
        pages = {59},
          doi = {10.3847/1538-4357/ab5c1b},
archivePrefix = {arXiv},
       eprint = {1911.11282},
 primaryClass = {astro-ph.EP},
       adsurl = {https://ui.adsabs.harvard.edu/abs/2020ApJ...889...59P}
}

@ARTICLE{1996ApJ...473..422P,
       author = {{Popham}, Robert and {Kenyon}, Scott and {Hartmann}, Lee and {Narayan}, Ramesh},
        title = "{Spectra and Line Profiles of FU Orionis Objects: Comparisons between Boundary Layer Models and Observations}",
      journal = {\apj},
         year = 1996,
        month = dec,
       volume = {473},
        pages = {422},
          doi = {10.1086/178155},
archivePrefix = {arXiv},
       eprint = {astro-ph/9606111},
 primaryClass = {astro-ph},
       adsurl = {https://ui.adsabs.harvard.edu/abs/1996ApJ...473..422P}
}

@ARTICLE{2007A&A...475...37S,
       author = {{Stamatellos}, D. and {Whitworth}, A.~P. and {Bisbas}, T. and {Goodwin}, S.},
        title = "{Radiative transfer and the energy equation in SPH simulations of star formation}",
      journal = {\aap},
         year = 2007,
        month = nov,
       volume = {475},
       number = {1},
        pages = {37-49},
          doi = {10.1051/0004-6361:20077373},
archivePrefix = {arXiv},
       eprint = {0705.0127},
 primaryClass = {astro-ph},
       adsurl = {https://ui.adsabs.harvard.edu/abs/2007A&A...475...37S}
}

@ARTICLE{2021ApJ...917...80S,
       author = {{Szab{\'o}}, Zs. M. and {K{\'o}sp{\'a}l}, {\'A}. and {{\'A}brah{\'a}m}, P. and {Park}, S. and {Siwak}, M. and {Green}, J.~D. and {Mo{\'o}r}, A. and {P{\'a}l}, A. and {Acosta-Pulido}, J.~A. and {Lee}, J.-E. and {Cseh}, B. and {Csoernyei}, G. and {Hanyecz}, O. and {Koenyves-T{\'o}th}, R. and {Krezinger}, M. and {Kriskovics}, L. and {Ordasi}, A. and {S{\'a}rneczky}, K. and {Seli}, B. and {Szak{\'a}ts}, R. and {Szing}, A. and {Vida}, K.},
        title = "{A Study of the Photometric and Spectroscopic Variations of the Prototypical FU Orionis-type Star V1057 Cyg}",
      journal = {\apj},
         year = 2021,
        month = aug,
       volume = {917},
       number = {2},
          eid = {80},
        pages = {80},
          doi = {10.3847/1538-4357/ac04b3},
archivePrefix = {arXiv},
       eprint = {2105.10405},
 primaryClass = {astro-ph.SR},
       adsurl = {https://ui.adsabs.harvard.edu/abs/2021ApJ...917...80S}
}

@ARTICLE{2022A&A...667A..12V,
       author = {{Valdivia-Mena}, M.~T. and {Pineda}, J.~E. and {Segura-Cox}, D.~M. and {Caselli}, P. and {Neri}, R. and {L{\'o}pez-Sepulcre}, A. and {Cunningham}, N. and {Bouscasse}, L. and {Semenov}, D. and {Henning}, Th. and {Pi{\'e}tu}, V. and {Chapillon}, E. and {Dutrey}, A. and {Fuente}, A. and {Guilloteau}, S. and {Hsieh}, T.~H. and {Jim{\'e}nez-Serra}, I. and {Marino}, S. and {Maureira}, M.~J. and {Smirnov-Pinchukov}, G.~V. and {Tafalla}, M. and {Zhao}, B.},
        title = "{PRODIGE - envelope to disk with NOEMA. I. A 3000 au streamer feeding a Class I protostar}",
      journal = {\aap},
         year = 2022,
        month = nov,
       volume = {667},
          eid = {A12},
        pages = {A12},
          doi = {10.1051/0004-6361/202243310},
archivePrefix = {arXiv},
       eprint = {2208.01023},
 primaryClass = {astro-ph.GA},
       adsurl = {https://ui.adsabs.harvard.edu/abs/2022A&A...667A..12V}
}

@ARTICLE{2023A&A...669A.137H,
       author = {{Hsieh}, T.-H. and {Segura-Cox}, D.~M. and {Pineda}, J.~E. and {Caselli}, P. and {Bouscasse}, L. and {Neri}, R. and {Lopez-Sepulcre}, A. and {Valdivia-Mena}, M.~T. and {Maureira}, M.~J. and {Henning}, Th. and {Smirnov-Pinchukov}, G.~V. and {Semenov}, D. and {Moller}, Th. and {Cunningham}, N. and {Fuente}, A. and {Marino}, S. and {Dutrey}, A. and {Tafalla}, M. and {Chapillon}, E. and {Ceccarelli}, C. and {Zhao}, B.},
        title = "{PRODIGE - envelope to disk with NOEMA. II. Small-scale temperature structure and streamer feeding the SVS13A protobinary based on CH$_{3}$CN and DCN}",
      journal = {\aap},
         year = 2023,
        month = jan,
       volume = {669},
          eid = {A137},
        pages = {A137},
          doi = {10.1051/0004-6361/202244183},
archivePrefix = {arXiv},
       eprint = {2211.05022},
 primaryClass = {astro-ph.GA},
       adsurl = {https://ui.adsabs.harvard.edu/abs/2023A&A...669A.137H}
}

@ARTICLE{1954ZA.....35...74W,
       author = {{Wachmann}, A.},
        title = "{Das bisherige Verhalten von FU Orionis. Mit 7 Textabbildungen}",
      journal = {\zap},
         year = 1954,
        month = jan,
       volume = {35},
        pages = {74},
       adsurl = {https://ui.adsabs.harvard.edu/abs/1954ZA.....35...74W}
}

@ARTICLE{2011ARA&A..49...67W,
       author = {{Williams}, Jonathan P. and {Cieza}, Lucas A.},
        title = "{Protoplanetary Disks and Their Evolution}",
      journal = {\araa},
         year = 2011,
        month = sep,
       volume = {49},
       number = {1},
        pages = {67-117},
          doi = {10.1146/annurev-astro-081710-102548},
archivePrefix = {arXiv},
       eprint = {1103.0556},
 primaryClass = {astro-ph.GA},
       adsurl = {https://ui.adsabs.harvard.edu/abs/2011ARA&A..49...67W}
}

@ARTICLE{2013MNRAS.435.2610N,
       author = {{Nelson}, Richard P. and {Gressel}, Oliver and {Umurhan}, Orkan M.},
        title = "{Linear and non-linear evolution of the vertical shear instability in accretion discs}",
      journal = {\mnras},
         year = 2013,
        month = nov,
       volume = {435},
       number = {3},
        pages = {2610-2632},
          doi = {10.1093/mnras/stt1475},
archivePrefix = {arXiv},
       eprint = {1209.2753},
 primaryClass = {astro-ph.EP},
       adsurl = {https://ui.adsabs.harvard.edu/abs/2013MNRAS.435.2610N}
}

@ARTICLE{2014AJ....147..140G,
       author = {{Gramajo}, Luciana V. and {Rod{\'o}n}, Javier A. and {G{\'o}mez}, Mercedes},
        title = "{Spectral Energy Distribution Analysis of Class I and Class II FU Orionis Stars}",
      journal = {\aj},
         year = 2014,
        month = jun,
       volume = {147},
       number = {6},
          eid = {140},
        pages = {140},
          doi = {10.1088/0004-6256/147/6/140},
archivePrefix = {arXiv},
       eprint = {1402.3240},
 primaryClass = {astro-ph.SR},
       adsurl = {https://ui.adsabs.harvard.edu/abs/2014AJ....147..140G}
}

\end{document}